\documentclass[aps,prl,showpacs,twocolumn,groupedaddress]{revtex4}

\usepackage{graphicx,bbm,amsthm, amsmath, amssymb}
\usepackage{color}

\begin{document}
\title{Chirality of triangular antiferromagnetic clusters as a qubit}

\author{B.~Georgeot$^{1,2}$ and F.~Mila$^{3}$}

\affiliation{$^1$  Universit\' e de Toulouse; UPS; Laboratoire de Physique
 Th\'eorique (IRSAMC);  F-31062 Toulouse, France \\
$^2$ CNRS; LPT (IRSAMC); F-31062 Toulouse, France \\
$^3$ Institute of Theoretical Physics, \'Ecole Polytechnique F\'ed\'erale de Lausanne,
1015 Lausanne, Switzerland}

\date{February 26, 2009}

\begin{abstract}
We show that the chirality of triangular antiferromagnetic clusters can
be used as a qubit even if it is entirely decoupled from the total spin of
the cluster. In particular, we estimate the orbital moment associated to
the chirality, and we show that it can be large enough to allow a direct
measurement of the chirality with a field perpendicular to the cluster.
Consequences for molecular magnets are discussed, and an alternative
implementation with Cu atoms on a surface is proposed, for which
one- and two-qubit gates are worked out in detail. Decoherence effects are also discussed.
\end{abstract}

\pacs{75.50.Xx, 03.67.Lx, 75.75.+a, 03.67.Pp}
\maketitle


It is widely recognized that the use of quantum properties of matter to store and
treat information could bring exciting new possibilities
(see e.g. \cite{nielsen}
and references therein).  It is therefore important
to find simple quantum systems (such as
qubits, two-level systems) on which some essential basic operations could be made reliably.
Requirements include the possibility to prepare and measure the system
in a specific basis, and also to couple subsystems
to perform coherent transformations of at least two qubits.
At the same time, it should be free enough from decoherence to allow
significant numbers of local operations to be performed, and
should also be scalable to allow for large arrays of qubits.
These requirements are very hard to fulfill, and many systems
lack at least one of them.  Solid state implementations have
the potentiality to be scalable to large size, and have been actively
studied, using e.g. Josephson junctions \cite{JJ}, electronic spin
in quantum dots \cite{Loss,DV}, nuclear spin chains \cite{berman}
or single-molecule magnets \cite{ring}.

In that respect, antiferromagnetic triangular clusters are particularly
interesting. First of all, a number of experimental realizations are already
available. But more importantly, the presence of an extra degree
of freedom in addition to the total spin, the chirality,
opens new possibilities with respect to purely magnetic qubits.
A first realization has been explored recently in \cite{Losstriangle}.
There it was
shown that the chirality induces a spin-electric effect whereby the
total spin of the cluster can be rotated by an external electric field
when spin-orbit interaction is taken into account. In that scheme, the qubit is
still essentially magnetic, and the measurement relies on the spin
magnetic moment of the qubit.

In this Letter, we discuss the alternative possibility of
qubits  entirely based on the chirality.
Getting rid of the spin is of course
a major advantage regarding decoherence since such a qubit would
be less sensitive to magnetic noise (see below), but this creates at the
same time a potential problem for measurement since there is no Zeeman
coupling of the qubit to an external magnetic field any more. As we show,
the solution to this problem relies on the presence of an orbital moment
associated to the chirality, as recently pointed out in another context
in \cite{Khomskii}. This allows the qubit to be measured
as if it had a spin, with the important difference however that a
magnetic field cannot induce a rotation of the qubit.
Note that a chiral doublet ground state has been
recently reported in a Dysprosium based molecular magnet \cite{Sessoli}.
For reasons explained below, we will discuss in detail an alternative
implementation with Cu atoms on a surface.

We note that in the quantum information community, the coding of qubits
in specific degrees of freedom has been investigated in the context
of noiseless subsystems \cite{viola}.
In particular, it has been shown that one can
use certain degrees of freedom of three spins as a qubit, which will be
isolated from certain types of noise \cite{filippo}.
In this regard, our proposal can be seen as a particularly
simple realization of a noiseless subsystem, where the protected degree of
freedom is identified with the physical chirality and the logical qubit is
directly manipulated and measured without any encoding procedure.

\begin{figure}[hbt]
\begin{center}
\includegraphics[width=.85\linewidth,angle=-00]{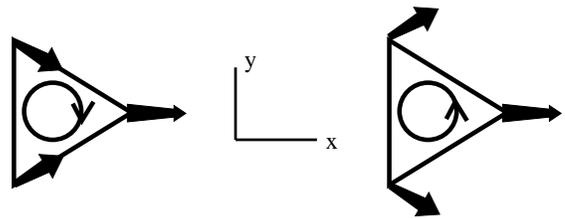}
\end{center}
\caption{Schematic view of the two
chiral states.  The two chiralities are illustrated by their classical analogs, namely
spin configurations forming umbrella-like structures with a net moment along the x-axis, and
with opposite chiralities. They are associated with currents circulating in opposite directions.
}
\label{triangle}
\end{figure}

The usual starting point to discuss triangular antiferromagnets is the Heisenberg model:
\begin{equation}
H_{Heis}=J_1 \vec{S_1}.\vec{S_2} + J_2 \vec{S_2}.\vec{S_3}+J_3\vec{S_3}.\vec{S_1}-g\mu_B \vec H. \vec S
\label{Hamil}
\end{equation}
where $\vec S$ is the total spin.
When $J_1=J_2=J_3=J$, the ground state is fourfold degenerate, and a convenient
basis is provided by the simultaneous eigenstates of the scalar
chirality $\vec{S_1}.(\vec{S_2}\times \vec{S_3})$ and of the
projection $S^\alpha$ of the total spin  in an arbitrary direction
$\alpha$:
\begin{eqnarray}
|R, \sigma\rangle & = & (|-\sigma \sigma \sigma\rangle
+\omega |\sigma -\sigma \sigma\rangle + \omega^2
|\sigma \sigma -\sigma\rangle)/\sqrt{3} \nonumber \\
|L, \sigma\rangle & = & (|-\sigma \sigma \sigma\rangle
+\omega^2 |\sigma -\sigma \sigma\rangle + \omega
|\sigma \sigma -\sigma \rangle)/\sqrt{3} \nonumber
\end{eqnarray}
where $\omega=\exp(2i\pi/3)$. $\sigma=\pm 1/2$ refers to $S^\alpha$
while $L$ and $R$ stand for
left and right and refer to the chirality. To get a twofold qubit, we apply a positive magnetic
field $H$ in the direction $\alpha$, which leaves us with a twofold degenerate ground state
$\{|R,+1/2\rangle, |L,+1/2\rangle\}$ separated from the other states by an energy $g\mu_B H$.
The two ground states are only distinguished by the chirality and thus constitute a
non-spin qubit (see Fig.(\ref{triangle}) for a pictorial description
of these states in the case where the magnetic field is in the $x$ direction).
This is different from the situation where non SU(2) invariant
terms in the Hamiltonian such as Dzyaloshinskii-Moriya
interactions lift the degeneracy, in which case the two ground states differ by the chirality
and by the spin \cite{Losstriangle}.

\begin{figure}[hbt]
\begin{center}
\includegraphics[width=.85\linewidth,angle=-00]{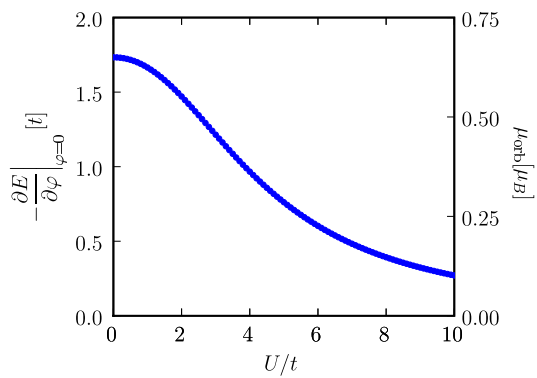}
\end{center}
\caption{(Color online) Orbital moment associated to the chirality as a function of the ratio $U/t$. Right scale:
in units of Bohr magneton, assuming the empirical dependence $t ({\rm eV}) = 10.67 /a({\rm \AA})^2$
for Cu $4s$ orbitals, where $a$ is the inter-site distance. Left scale: Opposite of the derivative of the
energy with the phase calculated at $\varphi=0$ in units of $t$.}
\label{orbital_moment}
\end{figure}

 Let us first concentrate on the measurement problem. Since we have
already used the magnetic field to create the qubit, it might seem hopeless to rely on the magnetic
field to measure it. This is not the case however. Indeed, let us turn to a more
microscopic description of the system in terms of a three-site Hubbard model:
\begin{equation}
\label{hubbard}
H_{Hub}=-t \sum_{i=1}^3 \sum_\sigma(c^\dagger_{i\sigma} c^{\vphantom{\dagger}}_{i+1\sigma} ) + h.c. + U \sum_{i+1}^3 n_{i\uparrow}n_{i\downarrow}
\end{equation}
with implicit periodic boundary conditions. With three electrons, a canonical transformation
maps this Hamiltonian onto the Heisenberg model with
coupling $J=4t^2/U$ to second order in $t/U$. However, as recently emphasized
in \cite{Khomskii}, charge fluctuations are not completely suppressed, and an orbital current proportional to $\vec{S_1}.(\vec{S_2}\times \vec{S_3})$ can be obtained by inverting the canonical transformation. These orbital currents create an orbital moment perpendicular to the plane of the triangle, which can couple to an external magnetic field.
Let us look at the magnitude of this orbital moment. In \cite{Khomskii},
this was calculated in perturbation theory and it was
found that it is of order $t^3/U^2$, hence very small. However, having in mind atomic Cu, where the outer electron
resides in the extended 4s orbital, we have calculated the orbital moment for an arbitrary value of the
ratio $t/U$. The orbital moment is related to the magnetic field by $\mu_{orb}=-\partial E/\partial B|_{B=0}$. The magnetic
field is taken perpendicular to the plane of the triangle and is treated with the Peierls substitution as a phase: $t=\vert t \vert e^{i\varphi}$. The phase is such that its circulation is equal to the flux of the field, which
leads to $\varphi=(\pi/2\sqrt{3})B a^2/\phi_0$, where $a$ is the intersite distance and $\phi_0=hc/e$ is the flux quantum. Using all symmetries, the ground state energy can be obtained as:
\begin{equation}
E=\rho \cos\left[\frac{1}{3} \arccos(A(B+C \sin 3 \varphi))+\frac {2\pi}{3}\right]+\frac{2U}{3}
\end{equation}
with $\rho=\sqrt{4U^2/9+12t^2}$, $A=-9/\rho(U^2+27t^2)$, $B=2U^3/27$ and $C=6 \sqrt{3}t^3$.
For small and large $U/t$, this leads to the following asymptotic expressions for the derivative:
\begin{equation}
-\frac{\partial E}{\partial \varphi}=\left\{\begin{array}{cc}
\sqrt{3} t (1-U^2/6t^2), & U/t \ll 1 \\
& \\
18 \sqrt{3} t^3/U^2, & U/t\gg 1
\end{array}
\right.
\label{partial_derivative}
\end{equation}
The full dependence as a function of $U/t$ is depicted in Fig.(\ref{orbital_moment}). The moment is a relatively slowly decreasing function of $U/t$: at $U/t=5$, it is still equal to about half its value at $U=0$. To get an estimate of the actual
magnitude of the moment, we use Harrisson's empirical dependence of the hopping integral on the intersite distance \cite{harrisson}
relevant for the 4s orbitals of Cu: $t ({\rm eV}) = 10.67 /a({\rm \AA})^2$. This leads to an orbital moment $\mu_{orb}=0.65 \mu_B$ at $U=0$, a
surprisingly large value in view of the results of Ref.~\cite{Khomskii}.
For realistic values of $U/t$, the orbital moment is thus
expected to be a sizable fraction of a Bohr magneton (see Fig.(\ref{orbital_moment})).
So one can use a field {\em parallel to the plane of the triangle}
to create the twofold degenerate ground state, and measure the chirality by inducing a component of the field {\em perpendicular to the cluster} to lift the degeneracy.

This effect will presumably be hard to detect in currently available Cu$_3$ clusters
for two reasons: first of all, the orbital moment depends on the magnitude of the inter-site hopping, which will
be much smaller for $3d$ electrons than for $4s$ electrons, as assumed in the above estimate. Besides, if the qubit is of mixed character due to spin-orbit coupling, an external magnetic field will couple to both the orbital moment
and the total spin, and the spin moment will dominate the splitting.

\begin{figure}[hbt]
\begin{center}
\includegraphics[width=.99\linewidth,angle=-00]{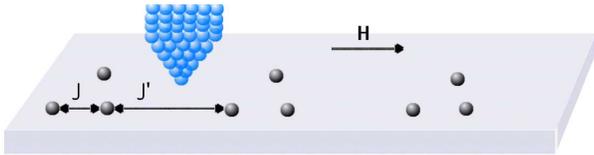}
\end{center}
\caption{(Color online) An example of chirality-based qubit chain:
  triangular clusters of Cu atoms deposited on a surface.
  $J$ is the interaction strength between spins
inside one triangle. Its spatial modulation induced by approaching an STM tip
of a triangle controls
the one-qubit gates (see text).  $J'$ is the interaction strength between closest
spins of two neighboring triangles induced by approaching an STM tip. It controls
the two-qubit gates (see text). $H$ is the in-plane magnetic field.
}
\label{layer}
\end{figure}

We thus turn to a description of an alternative implementation of such a
qubit (see Fig.(\ref{layer}) for a representation
of a chain of such qubits). The basic idea is to
work with atomic Cu deposited on a surface rather than with molecules with
Cu$^{2+}$ ions. The main advantages are: i) The absence
of spin-orbit coupling since the electron that carries the spin is in a $4s$ orbital;
ii) The extension of the $4s$ orbitals, which leads to a larger ratio $t/U$.
The imposition of a magnetic field perpendicular to the plane of the
triangles will align the qubit towards a state
whose orbital moment is polarized in the direction of the field.
This allows to prepare the qubit in a given chirality.
Positioning of atoms in well-defined arrays with precision
below the Angstr\"om level has been routinely realized using
STM tips to move and place atoms (see e.g. \cite{Hla} for a review).
Equilateral triangular arrays are especially easy to build
taking advantage of the crystalline structure of the atoms of the substrate.

To use this system as a qubit, one should be able to manipulate
this degree of freedom.
To create one-qubit gates and modify the state of one qubit,
one can modify the couplings in a controlled way
by use of STM tips (in a way which has
some similarity with the ones proposed in \cite{Losstriangle,universality}).
The control can be achieved by calibrating the
system, performing repeated preliminary modifications of the
couplings and measuring the result.  If the couplings
are transformed from  $J_1=J_2=J_3=J$ to
$J_1=J+\delta J_1$, $J_2=J+\delta J_2$, $J_3=J+\delta J_3$, the states
$|L\rangle$ and $|R\rangle$ are not any more eigenstates of
(\ref{Hamil}).  The Hamiltonian (\ref{Hamil}) in the basis
$(|L\rangle, |R\rangle$ reads:
\begin{equation}
H=\left(\begin{array}{cc} -D-3J/4 &
V_1+iV_2 \\
V_1-iV_2
& -D-3J/4 \end{array} \right)
\label{oneq}
\end{equation}
where the real parameters $D$, $V_1$ and $V_2$ are defined by
$D=-(\delta J_1+\delta J_2+\delta J_3)/4$ and
$V_1+iV_2=(\omega \delta J_1 + \delta J_2 +\omega^2 \delta J_3)/2$.
The evolution operator through the perturbation
is therefore $U=\exp-i(D.I +V_1 \tau^x-V_2 \tau^y)t=
\exp(-iDt)\exp-i(V_1 \tau^x-V_2 \tau^y)t $, where
$I$ is the identity operator and  $\tau^x$ and $\tau^y$ are Pauli matrices
associated with the chirality.
By an appropriate choice of the three parameters
$\delta J_1$, $\delta J_2$, $\delta J_3$ and of the time of evolution $t$,
it is therefore
possible to engineer any rotation of the qubit of axis $x$ or $y$.
It is known that any one-qubit gate can be written \cite{nielsen} as
$\exp(i\alpha) R_y(\beta) R_x (\gamma) R_y(\delta)$,
where $R_y(\theta)$ (resp. $R_x(\theta)$) is the rotation
of angle $\theta$ around the $y$-axis (resp. $x$-axis),
and $\alpha, \beta, \gamma, \delta$ are real parameters.  Thus
one can generate any one-qubit gate by a sequence
of well chosen modifications of the couplings inside the triangle.

To perform a two-qubit gate, one should couple two neighboring
triangles. If the triangles are sufficiently far apart, the
permanent coupling between the spins of different
triangles can be made small enough to be negligible. However, if there are
not too far apart, a STM tip approached between the closest spins
of two neighboring triangles can mediate an interaction between
two spins.  There is obviously an optimal distance between
the triangles to respect in the best possible way these two
requirements.  In the presence of the STM tip, the new interaction
$J'$ between the closest spins of each triangle creates an interaction
between the two triangles $i$ and $j$ which,
in the basis of the qubits,
takes the form
\begin{equation}
H'=(J'/9)\vec{{S}_i}.\vec{{S}_j}(1+2\tau_i^x+2\sqrt{3}\tau_i^y)(1- 4\tau_j^x),
\label{twoq}
\end{equation}
where $\vec{{S}_i}$ and $\vec{{S}_j}$ are the total spins of
the triangles \cite{kagome}.
Since the spins are polarized by the magnetic field parallel to the surface,
we can consider $\vec{{S}_i}.\vec{{S}_j}=1/4$ to be constant.
Since an arbitrary one-qubit gate can be made on each triangle,
this can be transformed
by rotations of the qubits to
$(J'/36)(1-4\tau_i^z)(1-4 \tau_j^z)$.
The evolution operator through this interaction is up to one-qubit gates
$\exp(-i(4J'/9)\tau_i^z\tau_j^z t)$; if applied for a time $t=9\pi/(16J')$
it generates the two-qubit gate
$\mbox{exp}(-i\pi/4) \mbox{diag}(1,i,i,1)$ in the basis $(|00\rangle,
|01\rangle,|10\rangle,|11\rangle)$,
equivalent up to
one-qubit operations to a CNOT gate \cite{siewert}.
Since the CNOT gate plus one-qubit gates form a universal set \cite{DV2},
the model proposed can generate any arbitrary sequence of gates
of a quantum algorithm.

The interaction time necessary to implement the one- and two-qubit
gates is $t_g \propto 1/\delta J$, where $\delta J$ is the
typical magnitude of the process ($\vert \omega \delta J_1 + \delta J_2 + \omega^2\delta J_3 \vert$
in Eq.(\ref{oneq}), $\delta J'$ in Eq.(\ref{twoq})).
If the tip is mechanically moved, time scales reached in experiments
of the early 90's were of the order of 10 ns \cite{steeves}.
This time can be significantly reduced, down to the picosecond range,
by using a non-moving photosensible tip optically addressed \cite{donati}.
The range $10\ {\rm ps} < t_g < 10\ {\rm ns}$ corresponds to $0.004\ {\rm K} < \delta J < 4\ {\rm K}$.
With a coupling $J$ of the order of 100 K, typical for kinetic exchange,
this corresponds to small, hence reasonable, modifications of the exchange integrals
that can be induced by the influence of the tip.
Note that the three important steps in the manipulation of a qubit
can be optimized independently since they depend on different
parameters: the orbital moment used to initialize and measure the qubit
depends on $t$ and $U$ in \eqref{hubbard} and is thus related to $J$ in \eqref{Hamil}, one-qubit gates depend on the modulations $\delta J_i$, and two-qubit gates are controlled
by the coupling $J'$ induced between the qubits.
This should help to tune the system to the optimal working point.
We also note that the system operated this way should be scalable
when increasing
the number of qubits: indeed, the qubits are individually addressed
and well-separated spatially.

Finally, let us discuss the problem of decoherence and why we
think such a qubit is well protected. There are a priori
two sources of decoherence: local vibrations, and magnetic noise.
Local vibrations can be a source of decoherence because they
change the bond lengths inside the Cu$_3$ cluster, hence the exchange integrals.
As usual, zero-point vibrations are not a source of decoherence. In
the present case, this is easy to see since, as long as the cluster has
a C$_3$ axis, the ground state for a given total spin is two-fold degenerate,
even if spin-phonon coupling is included. Indeed the degeneracy of the ground state on
which the qubit is based comes ultimately from the fact that the
C$_3$ group has a two-dimensional irreducible representation to which
the ground state belongs, and this will remain true for the system
including phonons.

Regarding thermal fluctuations, the main mechanism of decoherence comes from
the vibrations of individual Cu atoms in the potential well in which they
are located. The depth of the well varies from one system to the other, but
typical values of the energy barriers for diffusion are a fraction of an eV\cite{brune}.
The corresponding frequencies of harmonic vibrations are in the terahertz range,
i.e. 50 K. So, by working at low enough temperature, these thermal vibrations
can be exponentially suppressed. Since subkelvin temperatures are now accessible
to STM experiments, it should be possible to reduce this
source of decoherence very efficiently. The other mechanism is related to
vibrations of the substrate, or phonons. We believe that our qubit is naturally
protected against these vibrations. Indeed, at low temperature, only long-wave
length acoustic modes can be excited, and they are essentially decoupled
from chirality since it is only sensitive to
local differences in bond lengths inside the Cu$_3$ cluster.

The other source of decoherence is magnetic noise. The specificity
of the qubit proposed in the present paper is that the chirality is not coupled
to a uniform field, as stated before. So decoherence can only be induced by fluctuating
fields that are inhomogeneous on the scale of the Cu$_3$ cluster. A simple
calculation similar to the derivation of Eq.(\ref{twoq}) shows that the transition
amplitude between the two chirality states is equal to
$-(1/3)(h_1^\alpha+\omega^2 h_2^\alpha+\omega h_3^\alpha)$, where $\alpha$
is the direction of the field used to fix the orientation of the total
spin parallel to the plane, and $\vec h_i$ is the local field at site $i$ of the
cluster. Now, local spins such as nuclear spins in the neighborhood of
the cluster will produce inhomogeneous fields. The typical decoherence time
can be expected to be similar to that observed in nanomagnets, where such
a mechanism dominates, hence to be of the order of the microsecond \cite{Ardavan}. However,
by choosing a substrate with little or no nuclear spins, one could get rid
of this effect altogether. Interesting possibilities are silicon based
substrates such as silicon itself \cite{silicon} or SiO$_2$, where the only nuclear spins
are carried by $^{29}$Si, of natural abundance 4.7 \%, or carbon based
substrates such as graphene, with nuclear spins on $^{13}$C only, of natural
abundance 1 \%. A few percent of nuclear spins is probably acceptable since
spins far from the cluster will produce an almost homogeneous field, but in
any case isotopic purification could be used to reduce further this channel
of decoherence.

So, working at low enough temperature and with the appropriate substrate
should allow to reach decoherence times of several microseconds, much
longer than the operation time, which is in the nanosecond-picosecond range.

In conclusion, we have shown how to build a non-spin qubit
out of a magnetic cluster.  The two states of a qubit are
the lowest energy
states of opposite chirality of an equilateral triangular cluster of atoms with an in-plane
magnetic field. We have shown that electric fields generated by
STM tips are sufficient to create a universal set of quantum gates.
The presence of an orbital magnetic moment perpendicular to the plane of the triangle
enables to prepare the system and then to measure it at the end.
The structure of the chiral states makes them robust against noise, and decoherence times should be long enough to
perform many gates using current technology.
We thus think our proposal has several attractive features for
qubit implementation.

We thank the French ANR (project INFOSYSQQ),
the IST-FET program of the EC (project EUROSQIP), and the R\'egion Midi-Pyr\'en\'ees through
its Chaire d'Excellence Pierre de Fermat for funding, and S.~Gautier, S.~Rusponi and W.-D.~Schneider for discussions.


\end{document}